\documentclass{article}
\usepackage{spconf,amsmath,graphicx,bm,amssymb,color}

\newcommand{\be}{\mathbf{e}}

\newcommand{\bp}{\mathbf{p}}
\newcommand{\bq}{\mathbf{q}}

\newcommand{\bc}{\mathbf{c}}

\newcommand{\bI}{\mathbf{I}}

\newcommand{\bx}{\mathbf{x}}

\newcommand{\bs}{\mathbf{s}}

\newcommand{\bA}{\mathbf{A}}

\renewcommand{\frac}{\dfrac}

\newcommand{\SI}{{\mbox{SINR}}}
\newcommand{\T}{{{T}}}

\newcommand{\K}{\cal K}

\newcommand{\wuhao}{\fontsize{9pt}{\baselineskip}\selectfont}

\newtheorem{dingli}{Theorem~}[section]

\title{Joint Power and Admission Control Via $p$ Norm Minimization Deflation}
\name{Ya-Feng~Liu and Yu-Hong~Dai}
\address{State Key Lab.\ of Scientific and Engineering Computing\\
    Chinese Academy of Sciences\\
    Beijing 100190, China\\\texttt{\{yafliu,dyh\}@lsec.cc.ac.cn}}

\begin{document}
%
\maketitle
\wuhao {
\begin{abstract}
In an interference network, joint power and admission control aims
to support a maximum number of links
 at their specified signal to interference plus noise ratio (SINR) targets while using a minimum total transmission power.  
 In our previous work, we formulated the joint control problem as a sparse $\ell_0$-minimization problem and relaxed it to a $\ell_1$-minimization problem. In this work, we propose to {approximate} the $\ell_0$-optimization problem by a $p$ norm minimization problem where $0<p<1$, since intuitively $p$ norm will approximate $0$ norm better than $1$ norm. We first show that the $\ell_p$-minimization problem is strongly NP-hard and then derive a reformulation of it such that the well developed interior-point algorithms can be applied to solve it.
 The solution to the $\ell_p$-minimization problem can efficiently guide the link's removals (deflation). Numerical simulations show the proposed heuristic outperforms the existing algorithms.
\end{abstract}
%
\section{Introduction}
\label{sec:intro}
{Power control} is an effective tool for interference management in
cellular, ad-hoc, and cognitive underlay networks
\cite{convex_approximation,region,removals,Simple,ex3,msp,distributed,robust}.
The prevailing formulation of power control aims to use a minimum
total transmission power to support all links in an interference
network at their desired SINR targets. 
A longstanding issue associated with power control is that the
problem often becomes \emph{infeasible}, i.e., it is not possible to
simultaneously support all links in the network at their SINR
targets. In this case, we must adopt a joint power and admission
control approach to \emph{selectively} remove some links from the network so that the
remaining ones can be simultaneously supported at their desired SINR
levels. Our goal is to maximize the number of simultaneously
supportable links at their required SINR targets while using a minimum total
transmission power.

Theoretically, the joint power and admission control problem is
known to be NP-hard to solve to global optimality \cite{convex_approximation,removals} and to approximate to constant ratio global optimality \cite{msp}, so various
heuristic
algorithms \cite{convex_approximation,region,removals,msp,distributed,robust}
have been proposed for this
problem. Among them, the reference~\cite{convex_approximation} proposed a convex approximation-based algorithm for the joint power and admission control
problem. Instead of directly solving
the original NP-hard problem, the basic idea of the proposed linear
programming deflation (LPD) algorithm in \cite{convex_approximation}
is to approximate the problem by an appropriate convex problem. The
solution to the approximation problem~can be used to check the feasibility of the original problem and
guide link's
removals. {The removal procedure is terminated until all the remaining links in
the network are simultaneously supportable. 
The recent work~\cite{msp} developed another LP approximation-based new linear programming deflation (NLPD) algorithm for the joint power and admission control problem. In \cite{msp}, the joint power and admission control problem is first equivalently reformulated as a sparse $\ell_0$-minimization problem and then its $\ell_1$-convex approximation is used to derive a LP, which is different from the one in \cite{convex_approximation}. Again, the solution to the derived LP can guide an iterative link removal procedure, and the removal procedure is terminated if all the remaining links in
the network are simultaneously supportable.} 

Based on the sparse $\ell_0$-minimization reformulation in \cite{msp}, this paper proposes a new deflation algorithm based on $p$ ($0<p<1$) norm minimization for the joint power and admission control
problem. Compared to the $\ell_1$-minimization problem, the $p$ norm minimization problem is closer to the original $\ell_0$-optimization problem.
The $\ell_p$-approximation problem is solved by applying the efficient interior-point algorithm in \cite{ye} to solve its equivalent reformulation.
Numerical results show that the proposed algorithm compares favorably with the
existing approaches \cite{convex_approximation,region,msp} in
terms of the number of supported links, the total transmission power, and the CPU time.

{\emph{Notations:} We adopt the following notations in this paper. We denote the index
set $\{1,2,\cdots,K\}$ by ${\K}$. Lowercase boldface and uppercase
boldface are used for vectors and matrices, respectively. For a
given vector $\bx,$ the notations $\max\{\bx\},$ $[\bx]_k$ and $\|\bx\|_p:=\sum_{k}|[\bx]_k|^p~(0\leq p<1)\footnote{Strictly speaking, $\|\bx\|_p$ with $0\leq p<1$ is not a norm, since it does not satisfy the triangle inequality. However, we still call it $p$ norm for convenience in this paper.}$ stand
for its maximum entry, its $k$-th entry, and its $p$ norm, respectively. In particular, when $p=0,$ $\|\bx\|_0$ stands for the number of nonzero entries in $\bx.$ Finally, we use $\be$ to represent the vector of an appropriate size with all components
being one and $\bI$ to represent the identity matrix of an
appropriate size,
respectively.}

\section{Problem Formulation}
Consider a $K$-link (a link corresponds to a transmitter-receiver
pair) interference channel with channel gains $g_{kj}\geq0$ (from
transmitter $j$ to receiver $k$), noise power $\eta_k>0,$ SINR
target $\gamma_k>0,$ and power budget $\bar p_k>0$ for $k, j\in
{\K}:=\{1,2,\cdots,K\}$. Denote the power allocation vector by
$\bp=(p_1,p_2,\cdots,p_K)^\T$ and the power budget vector by $\bar
\bp=(\bar p_1,\bar p_2,\cdots,\bar p_K)^\T$. Treating interference
as noise, we can write the SINR at the $k$-th receiver as
\begin{equation}\nonumber\label{scale}\displaystyle
\SI_k=\frac{g_{kk}p_k}{\eta_k+\displaystyle\sum_{j\neq
k}g_{kj}p_j},\quad\forall~k\in\K.\end{equation}

The joint power and admission control problem can be mathematically
formulated as a two-stage optimization problem \cite{convex_approximation}. Specifically, the
first stage
maximizes the number of admitted links:
\begin{equation}\label{MSP1}
\begin{array}{ll}
\displaystyle \max_{{\bp,\,{\cal S}}} & \displaystyle |\cal S| \\
[5pt] \mbox{s.t.} & \displaystyle
\SI_k\geq \gamma_k,~k\in\cal S\subseteq\cal K,\\
&\bm{0} \leq \bp\leq \bar\bp.
\end{array}
\end{equation}
We use ${{\cal S}_0}$ to denote the optimal solution for problem \eqref{MSP1} and call it \emph{maximum admissible set}.
Notice that the solution for \eqref{MSP1}
might not be unique. The second stage minimizes the total
transmission power required to support the admitted links in ${{\cal
S}_0}$:
\begin{equation}\label{MSP2}
\begin{array}{cl}
 \displaystyle\min_{\left\{p_k\right\}_{k\in{\cal S}_0}} & \sum_{k\in {{\cal S}_0}} p_k \\
  \mbox{s.t.} &  \SI_k\geq \gamma_k,~k\in{{\cal S}_0},\\
&\displaystyle 0\leq p_k\leq \bar p_k,~
k\in{\cal S}_0.
\end{array}
\end{equation}
Due to the special choice of ${\cal S}_0,$ power control problem
\eqref{MSP2} is feasible and can be efficiently and distributively solved by the
Foschini-Miljanic algorithm \cite{Simple}.

\section{Review of the NLPD Algorithm}\label{secreview}

Since the developed algorithm in this paper follows the similar idea as the NLPD algorithm in \cite{msp}, we first briefly review the NLPD algorithm in this section.
 The basic idea of the NLPD algorithm is to
update the power and check whether all links can be supported or
not. If the answer is yes, then terminate the algorithm; else drop one link from the
network and update the power again. The above process is repeated
until all the remaining links are supported.

We begin with the introduction of an equivalent normalized channel on which the NLPD algorithm is
based. In particular, we use $\bq=\left(q_1,q_2,\cdots,q_K\right)^T$
with
$q_k={p_k}/{\bar p_k}$
~to denote the normalized power allocation vector, and use
${\bc}=\left(c_1,c_2,\cdots,c_K\right )^T$ with
  $c_k={\left(\gamma_k\eta_k\right)}/{\left(g_{kk}\bar p_k\right)}>0$~
to denote the normalized noise vector. 
We denote the normalized channel matrix by $\bA\in {\mathbb R}^{K\times
K}$ with the $(k,j)$-th entry
\begin{equation*}\label{A}
a_{kj}=\left\{\begin{array}{cl}
1,&\text{if~}k=j;\\
\displaystyle - \frac{\gamma_kg_{kj}\bar p_j}{g_{kk}\bar p_k},&\text{if~}k\neq j.
\end{array}
\right.
\end{equation*}
In fact, $|a_{kj}|$ is the normalized channel gain.
It is simple to check that 
$\SI_k\geq\gamma_k$ if and only if $[\bA\bq-\bc]_k \geq 0.$

In \cite{msp}, we reformulate the two-stage joint power and
admission control problem \eqref{MSP1} and \eqref{MSP2} as a
single-stage optimization problem
\begin{equation}\label{sparse2}
\begin{array}{cl}
\displaystyle \min_{\bq_e,\,\bq} & \|\bq_e\|_0+\alpha\bar\bp^T\bq \\
\mbox{s.t.} & \displaystyle \bq_e=\bc-\bA\bq,\\&\mathbf{0}\leq
\bq\leq \be,
\end{array}
\end{equation}
where $0< \alpha<\alpha_1:=1/{\be^T\bar\bp},$ {and $[\bq_e]_k$ measures the
excess transmission power \cite{convex_approximation} that the transmitter of link $k$ needs in the normalized channel in
order to be served with its desired SINR target (assuming all other
links keep their transmission powers unchanged).} 
Notice that the formulation \eqref{sparse2} is capable of finding the maximum admissible set with minimum total transmission power.
~Since problem \eqref{sparse2} is still NP-hard, we further consider
its $\ell_1$-convex {approximation} (equivalent to a LP)
\begin{equation}\label{ll1}
\begin{array}{cl}
\displaystyle \min_{\bq_e,\,\bq} & \|\bq_e\|_1+\alpha\bar\bp^T\bq\\
\mbox{s.t.} & \displaystyle \bq_e=\bc-\bA\bq,\\&\mathbf{0}\leq
\bq\leq \be.
\end{array}
\end{equation}
By solving \eqref{ll1}, we know whether
all links in the network can be simultaneously supported or not. 
If
not, we drop the link
\begin{equation}\label{remove2}k_0=\arg\max_{k\in\cal K}\left\{\sum_{j\neq k}\left(|a_{kj}|\left[\bq_e\right]_j+|a_{jk}|\left[\bq_e\right]_k\right)\right\}.\end{equation}

An easy-to-check necessary condition
\begin{equation}\label{necessary}
\left(\bm{\mu}^+\right)^T\be-\left(\bm{\mu}^-+\be\right)^T\bc\geq0
\end{equation}
 for all links in the
network to be simultaneously supported is also derived in
\cite{msp}, where $\bm{\mu}^+=\max\left\{\bm{\mu},\mathbf{0}\right\},$
$\bm{\mu}^-=\max\left\{-\bm{\mu},\mathbf{0}\right\},$ and
$\bm{\mu}=\bA^\T\be$. The necessary condition allows us to iteratively
remove strong interfering links from the network. In particular, we remove the link
$k_0$ according to the scheme
\begin{equation}\label{sum}
k_0=\arg\max_{k\in\cal K}\left\{\sum_{j\neq
k}|a_{kj}|+\sum_{j\neq k}|a_{jk}|+c_k\right\}
\end{equation} until \eqref{necessary} becomes true.

The NLPD algorithm can be described as
follows.\vspace{-0.5cm}\begin{center}
\framebox{
\begin{minipage}{8.3cm}
\flushright
\begin{minipage}{8.3cm}
\centerline{\bf The NLPD Algorithm}\vspace{0.05cm} \textbf{Step 1.}
Initialization: Input data
$\left(\bA,\bc,\bar\bp\right).$\\[2.5pt]
\textbf{Step 2.} Preprocessing: Remove link $k_0$ iteratively
according to \eqref{sum} until condition \eqref{necessary} holds
true.\\[2.5pt]
\textbf{Step 3.} Power control: Solve problem \eqref{ll1}; check
whether all links are supported: if yes, go to \textbf{Step
5}; else go to \textbf{Step 4}. \\[2.5pt]
\textbf{Step 4.} Admission control: Remove link  $k_0$ according to
\eqref{remove2}, set ${\K}={\K}/\left\{k_0\right\},$ and go to
\textbf{Step 3}.\\[2.5pt]
\textbf{Step 5.} Postprocessing: Check the removed links for possible admission.
\end{minipage}
\end{minipage}
}
\end{center}

\section{A $p$ Norm Minimization Deflation Algorithm}
\label{reformulation}
In this section, we develop a new deflation
algorithm based on $\ell_p$-minimization for the joint control problem \eqref{MSP1} and \eqref{MSP2}. As seen in Section \ref{secreview}, the original $\ell_0$-minimization problem \eqref{sparse2} is \emph{successively} approximated by the $\ell_1$-minimization problem \eqref{ll1} 
in the NLPD algorithm. Intuitively, the $p$ ($0<p<1$) norm minimization problem
\begin{equation}\label{lp}
\begin{array}{cl}
\displaystyle \min_{\bq_e,\,\bq} & \|\bq_e\|_p+\alpha\bar\bp^T\bq \\
\mbox{s.t.} & \displaystyle \bq_e=\bc-\bA\bq,\\&\mathbf{0}\leq
\bq\leq \be
\end{array}
\end{equation}
should approximate \eqref{sparse2} better than \eqref{ll1}. This is the motivation for the development of the new deflation algorithm based on the $p$ norm minimization for the joint power and admission control problem.

Comparing the $1$ norm minimization problem \eqref{ll1} and the $p$ norm minimization problem \eqref{lp}, we see that problem \eqref{ll1} is convex while problem \eqref{lp} is nonconvex (for its objective function is nonconvex with respect to $\bq_e$).
Generally speaking, convex problems are relatively easy to solve, while nonconvex
optimization problems are difficult to solve. However, not all nonconvex problems
are hard since the lack of convexity may be due to an inappropriate
formulation. In fact, many nonconvex optimization
problems admit a convex reformulation. Therefore, convexity
is useful but unreliable to test the computational intractability
of an optimization problem. A more robust tool is the computational
complexity theory \cite{Complexitybook}.
%

We show that problem \eqref{lp} is strongly NP-hard. The proof is based on a polynomial time reduction from the MAX-2UNANIMITY problem, which is shown to be strongly NP-hard in \cite{coordinated}.

\begin{dingli}\label{thm1}
  The $\ell_p$-minimization problem \eqref{lp} is strongly NP-hard if $0\leq p<1.$
\end{dingli}

The complexity result in Theorem \ref{thm1} motivates us to approximately solve problem \eqref{lp}.
~Next, we first give a reformulation of problem \eqref{lp}, and then propose to use the interior-point algorithm developed in
\cite{ye} to solve it.

\begin{dingli}\label{thm2}
The $\ell_p$-minimization problem \eqref{lp} can be equivalently reformulated as
\begin{equation}\label{llp}
\begin{array}{cl}
\displaystyle \min_{\bq_e,\,\bq} & \sum_{k}[\bq_e]_k^p +\alpha\bar\bp^T\bq \\
\mbox{s.t.} & \displaystyle \bq_e=\bc-\bA\bq,\\&\mathbf{0}\leq
\bq\leq \be,~\bq_e\geq\mathbf{0}.
\end{array}
\end{equation}
\end{dingli}
{A rigorous proof of Theorem \ref{thm2} shall be given in the journal version. Here we just shed some light on why it does not harm optimality to restrict $\bq_e\geq\mathbf{0}.$ Notice that problem \eqref{llp} is equivalent to
\begin{equation*}\label{restriction}
  \begin{array}{ll}
    \displaystyle \min_{\bq_e,\,\bq} & \|\bq_e\|_p+\alpha\,\bar\bp^T\bq\\
\mbox{s.t.} &\displaystyle \bq_e=\bc-\bA\bq,\\&\mathbf{0}\leq
\bq\leq \be,~\bq_e\geq\mathbf{0}.
  \end{array}
\end{equation*}
Thus, to show the equivalence of \eqref{lp} and \eqref{llp}, it suffices to show that
any optimal solution $(\tilde\bq_e,\tilde\bq)$ of \eqref{lp} always satisfies
$\tilde\bq_e=\bc-\bA\tilde\bq\geq\mathbf{0}$. In fact, assume the contrary that $|{\K^+}|\geq1,$ where $\K^+$$=\{\,k~|~[\tilde\bq_e]_k>0\,\}$,
$\K^=$$=\{\,k~|~[\tilde\bq_e]_k=0\,\}.$ Then by the Balancing Lemma (see \cite[Lemma 1]{msp}), we can appropriately reduce the power of links in $\K^+\cup\K^=$ so that both the first term and the second term in the objective of \eqref{lp} are strictly decreased.}

Based on Theorem \ref{thm2}, by introducing a slack variable $\bs\geq\bm{0}$ to problem \eqref{llp}, we see that problem \eqref{lp} is actually equivalent to
\begin{equation}\label{ye}
\begin{array}{cl}
\displaystyle \min_{\bq_e,\,\bq,\,\bs} & \sum_{k}[\bq_e]_k^p +\alpha\bar\bp^T\bq \\
\mbox{s.t.} & \displaystyle \bq_e=\bc-\bA\bq,\,\bs+\bq=\be\\&\bq\geq\mathbf{0},\,\bq_e\geq\mathbf{0},\,\bs\geq\mathbf{0}.
\end{array}
\end{equation}Now, we can apply the interior-point algorithm in \cite{ye} to solve problem \eqref{ye}. 
Similar to \cite{ye}, we can prove that the potential reduction interior-point algorithm returns an $\epsilon$-KKT \cite{ye2} or $\epsilon$-global solution of problem \eqref{ye} in no more than $O\left((\frac{3K}{\epsilon})\log(\frac{1}{\epsilon})\right)$ iterations.

One may ask why we wish to use interior-point algorithms to solve problem \eqref{ye}? The reasons are the following.
First, the objective function of problem \eqref{ye} is \emph{differentiable} in the interior feasible region. Moreover, we are actually interested in finding a sparse solution $\bq_e$ of problem \eqref{ye}; if we start from a solution, some of whose entries are already zero, then it is very hard to make it nonzero. 
In contrast, if we start from an interior point, the interior-point algorithm may generate a sequence of interior points that bypasses solutions with the wrong zero supporting set and converges to the true one. This is exactly the idea of the interior-point algorithm developed in \cite{ye2} for the nonconvex quadratic programming. 

The proposed $p$ norm minimization deflation (PNMD) algorithm is given as follows. There are two unclear points in the PNMD algorithm. One is how to compute the parameter $\alpha$ in problem \eqref{ye}, and the other is which removal strategy will be used in the admission control step. Next, we make clear of these two points, i.e., we shall use $\alpha$ given in \eqref{optalpha2} and the removal strategy \eqref{smart2} in the new deflation algorithm.

In the NLPD algorithm \cite{msp}, the parameter $\alpha$ is given by 
\begin{equation}\label{optalpha}\alpha=\left\{\!\!\!\!
  \begin{array}{rll}
  &c_1\alpha_1, & \text{if}~\rho(\bI-\bA)\geq 1,\\
  &c_2\min\left\{\alpha_1,\,\alpha_2\right\}, & \text{if}~\rho(\bI-\bA)< 1,
  \end{array}\right.
\end{equation}
where 
$0<c_1\leq c_2<1$ are two constants, and $\alpha_1$ is determined by the equivalence between problem \eqref{sparse2} and the joint problem \eqref{MSP1} and \eqref{MSP2}, and $\alpha_2$ is determined by the so-called ``Never-Over-Removal'' property. Since the $\ell_p$-minimization problem \eqref{lp} is closer to the $\ell_0$-minimization problem \eqref{sparse2}, we relax the parameter $\alpha$ in \eqref{optalpha} to 
\begin{equation}\label{optalpha2}\alpha=\left\{\!\!\!\!
  \begin{array}{rll}
  &c_1\alpha_1, & \text{if}~\rho(\bI-\bA)\geq 1,\\
  &\min\left\{c_2\alpha_1,\,c_3\alpha_2\right\}, & \text{if}~\rho(\bI-\bA)< 1,
  \end{array}\right.
\end{equation}where $c_3>c_2,$ $0<c_1, c_2<1$ are three constants.

Having obtained the solution $\left(\bq_e,\,\bq,\,\bs\right)$ of problem \eqref{ye}, we use the removal strategy called SMART rule in \cite{removals} to drop the link $k_0$ according to
\begin{equation}\label{smart2}
k_0=\arg\max_{k\in {\cal K}}\left\{\sum_{j\ne k
}|a_{kj}|q_j+\sum_{j\ne k}|a_{jk}|q_k+c_k\right\}.
\end{equation} The above operation can be interpreted as removing the
link with the largest interference plus noise footprint in the
\emph{normalized} network.

\vspace{-0.3cm}
\begin{center}
\framebox{
\begin{minipage}{8.3cm}
\flushright
\begin{minipage}{8.3cm}
\centerline{\bf The PNMD Algorithm}\vspace{0.05cm} \textbf{Step 1.}
Initialization: Input data
$\left(\bA,\bc,\bar \bp\right).$\\[2.5pt]
\textbf{Step 2.} Preprocessing: Remove link $k_0$ iteratively
according to \eqref{sum} until condition \eqref{necessary} holds
true.\\[2.5pt]
\textbf{Step 3.} Power control: Compute the parameter $\alpha$ and solve problem \eqref{ye}; check
whether all links are supported: if yes, go to \textbf{Step 5};
else
go to \textbf{Step 4}. \\[2.5pt]
\textbf{Step 4.} Admission control: Remove link  $k_0$ according to
some removal strategy, set ${\K}={\K}/\left\{k_0\right\},$ and go to
\textbf{Step 3}.\\[2.5pt]
\textbf{Step 5.} Postprocessing: Check the removed links for possible admission.
\end{minipage}
\end{minipage}
}
\end{center}

\section{Numerical Simulations}\label{nume}
We generate the same channel parameters as in \cite{convex_approximation} in our numerical simulations, i.e., each
transmitter's location obeys the uniform distribution over a $2$
Km~$\times$~$2$ Km square and the location of its corresponding
receiver is uniformly generated in a disc with radius $400$ m;
channel gains are given by $g_{kj}=1/d_{kj}^4~(\forall~k,\,j\in\K),$
where $d_{kj}$ is the Euclidean distance from the link of
transmitter $j$ to the link of receiver $k.$ Each link's SINR target
is set to be $\gamma_k=2~\text{dB}~(\forall~k\in\K)$ and the noise
power is set to be $\eta_k=-90~\text{dBm}~(\forall~k\in\K)$. The
power budget of the link of transmitter $k$ is
$\bar p_k=2p_k^{\min} ~(\forall~k\in\K),$  where $p_k^{\min}$ is
the minimum power needed for link $k$ to meet its SINR requirement
in the absence of any interference from other links.

\begin{figure}[!t]
     \centering
     \centerline{\includegraphics[width=8.2cm]{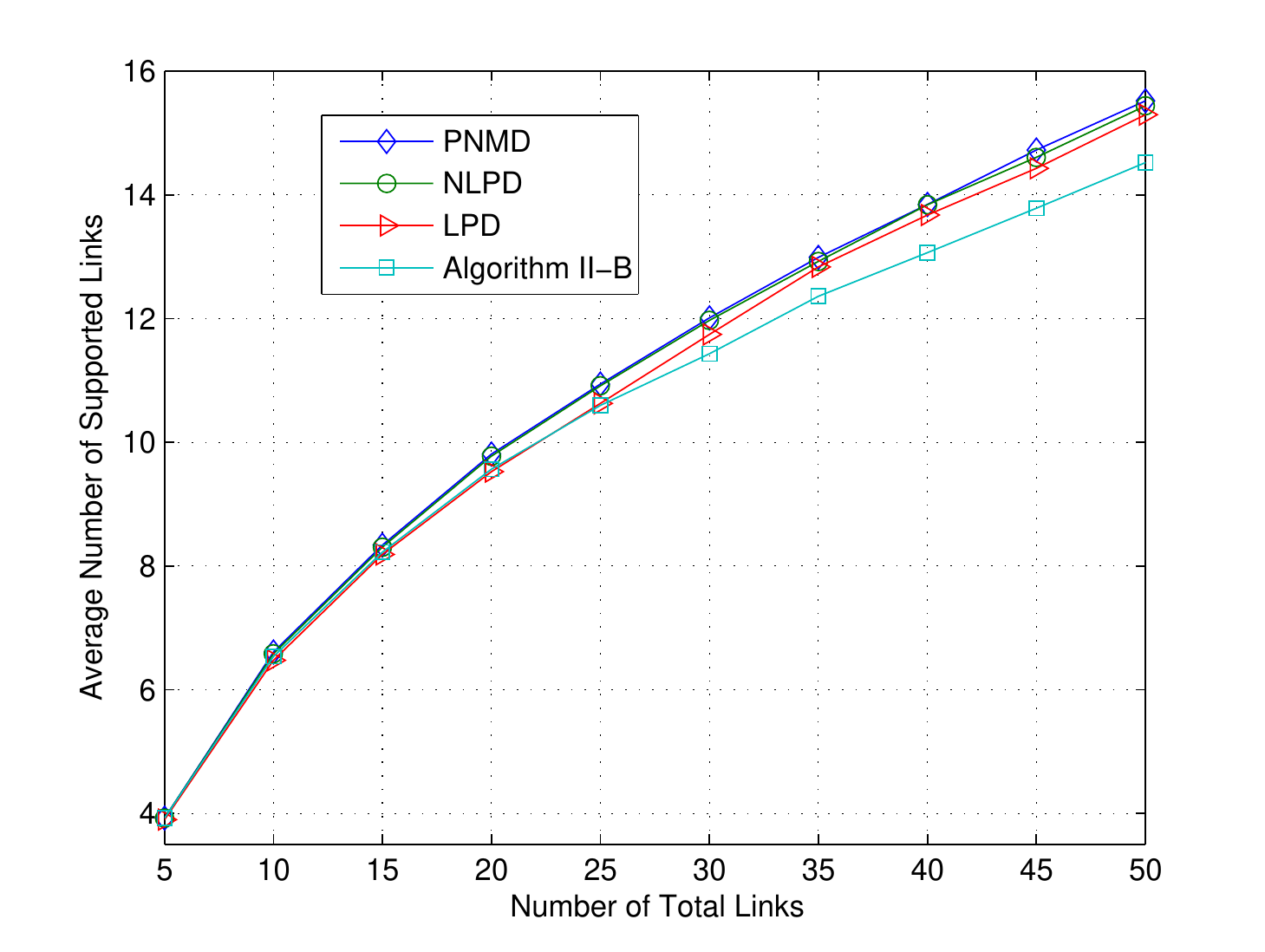}}\vspace{-0.3cm}
     \caption{\wuhao{Average number of supported links versus the number of total
     links.}}
     \label{user1}
     \end{figure}
\begin{figure}[!t]
     \centering
     \centerline{\includegraphics[width=8.2cm]{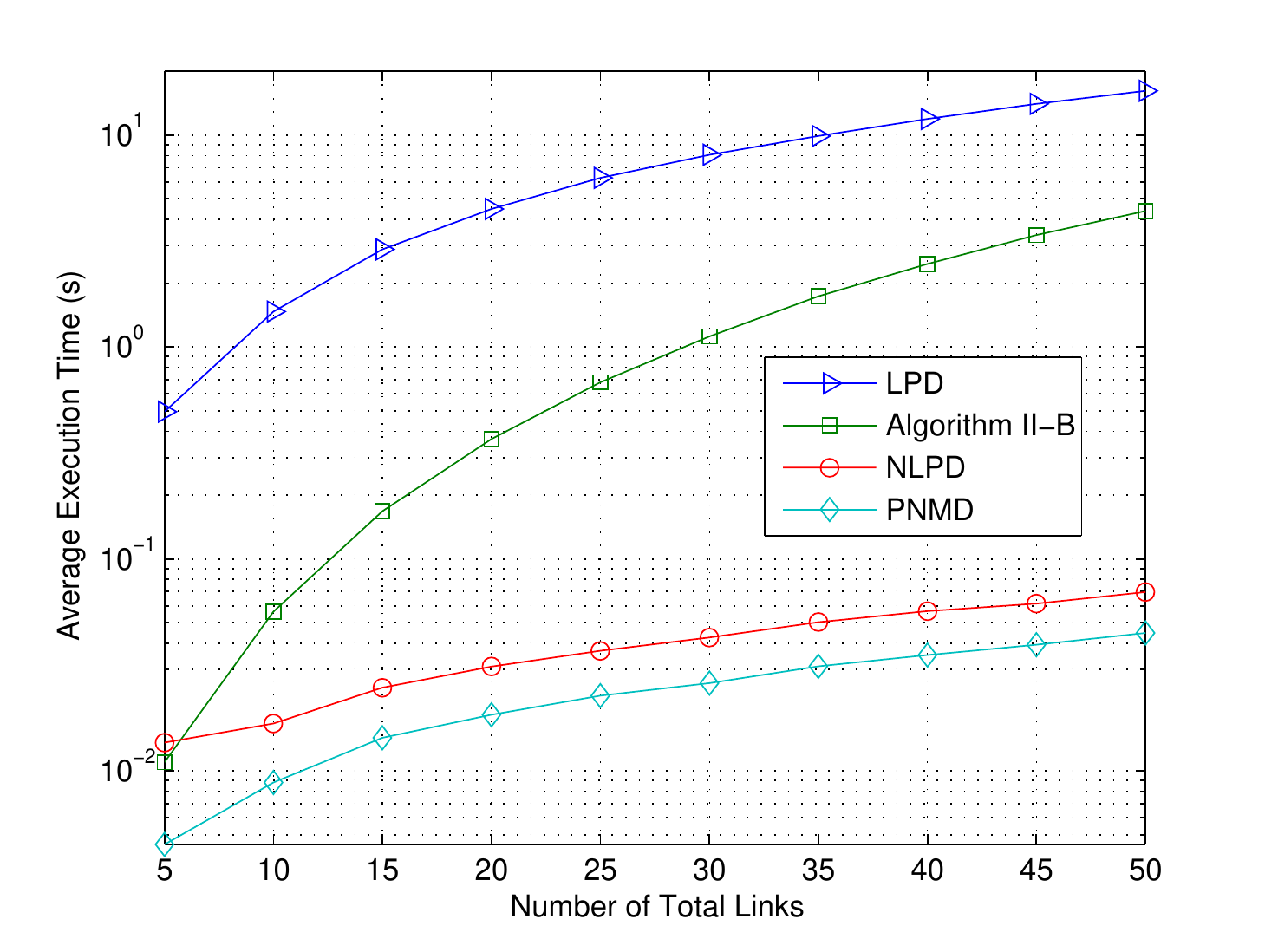}}\vspace{-0.3cm}
     \caption{\wuhao{Average CPU time versus the number of total
     links.}}
     \label{time1}
     \end{figure}
\begin{figure}[!t]
     \centering
     \centerline{\includegraphics[width=8.2cm]{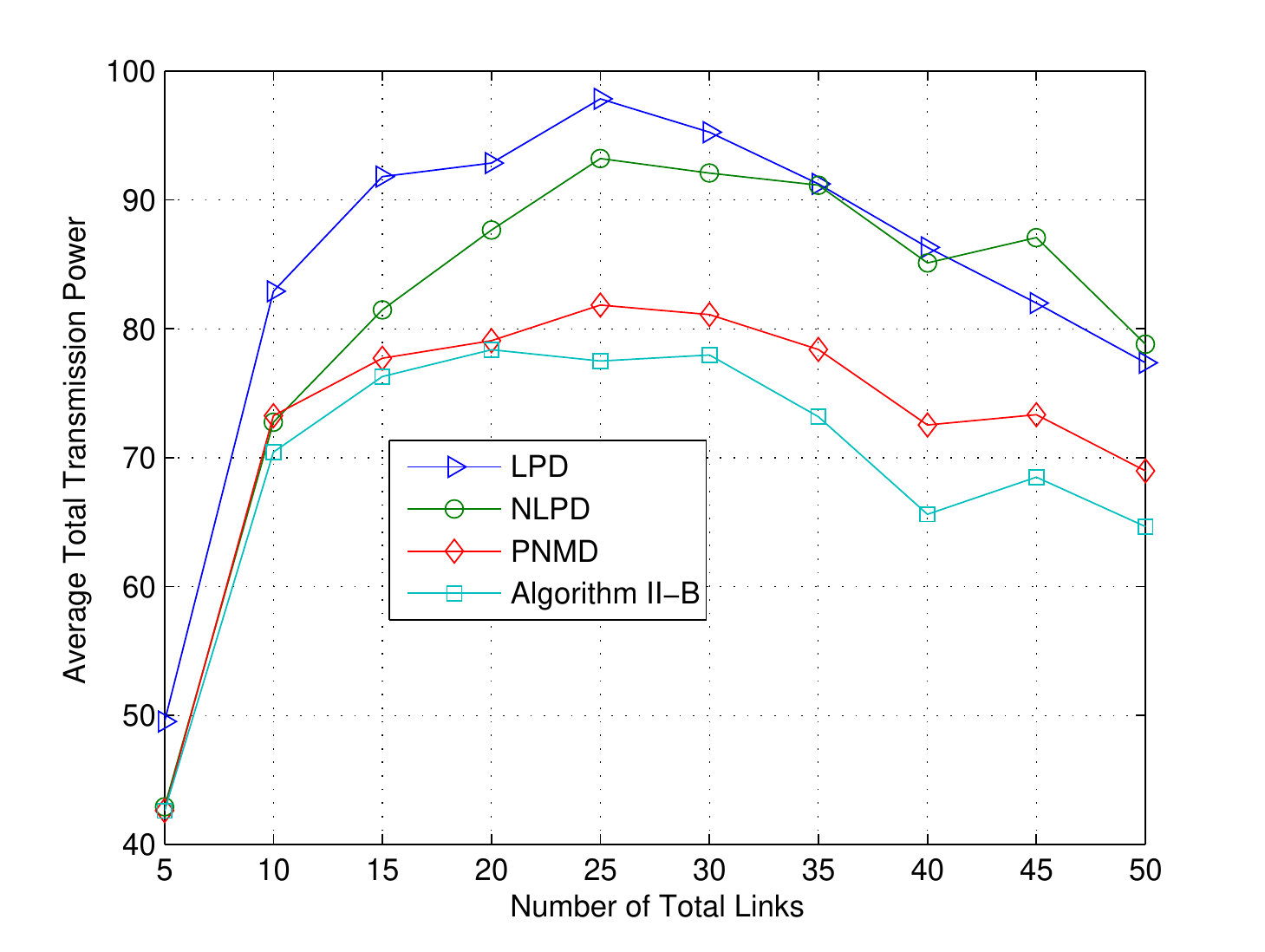}}\vspace{-0.3cm}
     \caption{\wuhao{Average transmission power versus the number of total
     links.}}
     \label{power1}
     \end{figure}

%
%
%

The parameter $p$ in problem \eqref{lp} is set to be $0.5$ and the ones in \eqref{optalpha2} are set to be $c_1=c_2=0.2$~and~$c_3=4$. The number of
supported links, the total transmission power, and the CPU time are the metrics we employ
to compare the performance of the
proposed PNMD algorithm with that of the LPD
algorithm in \cite{convex_approximation}, the Algorithm II-B in
\cite{region}, and the NLPD algorithm in \cite{msp}. All figures are obtained by averaging over $200$ Monte-Carlo runs.


%
Figs.\ \ref{user1}, \ref{time1}, and \ref{power1} indicate that the PNMD
algorithm can take less CPU time to support more links while with less total transmission power than the existing algorithms (except the Algorithm II-B).
As shown in Fig. \ref{power1}, the Algorithm II-B transmits the least power among the tested
algorithms. This is because the Algorithm II-B supports the least number of links; see Fig. \ref{user1}.
In particular, compared to the NLPD algorithm\footnote{To the best of our knowledge, the NLPD algorithm is so far the best removal-based algorithm for the joint power and admission control problem. It is shown in \cite{msp} that the NLPD algorithm can achieve more than 98\% of global optimality in terms of the number of supported links when $K\leq18.$}, the proposed PNMD algorithm can support (slightly) more links with much less total transmission power, and at the same time takes less CPU time.

{The performance improvement of the proposed PNMD algorithm over the NLPD algorithm is mainly attributed to the $\ell_p$-approximation problem \eqref{lp}.
The simulation results in Fig. \ref{user1} and Fig. \ref{power1} show that the admissible set ${\cal{S}}_1$ obtained by the proposed PNMD algorithm based on the $\ell_p$-approximation problem \eqref{lp} is ``better'' than the admissible set ${\cal{S}}_2$ obtained by the NLPD algorithm based on the $\ell_1$-approximation problem \eqref{ll1}, i.e., although the cardinality of the two admissible sets ${\cal{S}}_1$ and ${\cal{S}}_2$ is nearly equal to each other, it takes much less total transmission power to support the links in ${\cal{S}}_1$ than to support the links in ${\cal{S}}_2.$ This is consistent with our intuition that the $p~(0<p<1)$ norm minimization problem \eqref{lp} is capable of approximating the $\ell_0$-minimization problem \eqref{sparse2} better than the $\ell_1$-minimization problem \eqref{ll1}} and the fact that the maximum admissible set for the joint power and admission control problem may not be unique.


\section{Conclusions}
In this paper, we have developed a $p$ $(0<p<1)$ norm minimization deflation algorithm for the joint power and admission control problem.
Numerical simulations show the proposed algorithm outperforms state-of-the-arts in \cite{convex_approximation,region,msp} in terms of the number of supported links, the total transmission power, and the CPU time .
}

\begin{thebibliography}{1}\wuhao{
\bibitem{convex_approximation} I. Mitliagkas, N. D. Sidiropoulos,
and A. Swami, ``Joint power and admission control for ad-hoc and
cognitive underlay networks: Convex approximation and distributed
implementation,'' \emph{IEEE Trans. Wireless Commun.}, vol. 10, no.
12, pp. 4110--4121, Dec. 2011.
\bibitem{region} H.~Mahdavi-Doost, M.~Ebrahimi, and A.~K.~Khandani,~``Characterization of SINR region for interfering links with constrained power,'' \emph{IEEE Trans. Inf. Theory}, vol. 56, no.~6, pp. 2816--2828, Jun.
2010.
\bibitem{removals} M. Andersin, Z. Rosberg, and J. Zander, ``Gradual removals in cellular PCS with constrained power control and
noise,'' \emph{Wireless Netw.}, vol. 2, no. 1, pp.~27--43, Mar. 1996.
\bibitem{Simple} G.~J.~Foschini and Z. Miljanic, ``A simple distributed autonomous power control algorithm and its convergence,'' \emph{IEEE Trans.
Veh. Technol.}, vol. 42, no.~4, pp. 641--646, Nov. 1993.
\bibitem{ex3} S. A. Grandhi, J. Zander, and R. Yates, ``Constrained power control,'' \emph{Wireless
Personal Commun.}, vol. 1, no. 4, pp.~257--270, 1995.
\bibitem{msp} Y.-F. Liu, Y.-H. Dai, and Z.-Q. Luo, ``Joint power and admission control via linear programming deflation,'' {\emph{IEEE Trans. Signal Process.}, vol. 61, no. 6, pp. 1327--1338, Mar. 2013.}
{\bibitem{distributed} Y.-F. Liu, ``An efficient distributed joint power and admission control algorithm,'' in \emph{Proc. 31th Chinese Control Conference, Signal Processing and Optimization Session,} July, 2012, pp. 5508--5512.
\bibitem{robust} Y.-F. Liu and E. Song, ``Sample approximation-based deflation approaches for chance SINR constrained joint power and admission control.'' Available online: http://arxiv.org/abs/1302.5973.}
\bibitem{ye} D. Ge, X. Jiang, and Y. Ye, ``A note on the complexity of $L_p$ minimization,'' to appear in \emph{Math. Prog.} Available online: http://www.stanford.edu/~yyye/lpmin\_v14.pdf.
\bibitem{Complexitybook} M. R. Garey and D. S. Johnson, \emph{Computers and Intractability: A Guide to the Theory of
NP-Completeness}. SF, U.S.A.:~W. H. Freeman and Company, 1979.
\bibitem{coordinated} Y.-F. Liu, Y.-H. Dai, and Z.-Q. Luo,
``Coordinated beamforming for MISO interference channel: Complexity
analysis and efficient algorithms,'' \emph{IEEE Trans. Signal
Process.,} vol. 59, no. 3, pp. 1142--1157, Mar. 2011.
\bibitem{ye2} Y. Ye, ``On the complexity of approximating a KKT point of quadratic proramming,'' \emph{Math. Prog.,} vol. 80, pp. 195--211, 1998.


%
%








}
\end{thebibliography}
\end{document}